\documentclass[vecphys]{svmult}

\usepackage{makeidx}         
\usepackage{graphicx}        
\usepackage{multicol}        
\usepackage[bottom]{footmisc}

\makeindex             

\newcommand\ion[2]{#1$\,${{\sc #2}}}

\begin{document}

\title*{\ion{Ca}{ii} and \ion{Na}{i} absorption signatures from the circumgalactic gas of the Milky Way}
\author{N. Ben Bekhti\inst{1} \and M. Murphy\inst{2} \and P. Richter\inst{3} \and T. Westmeier\inst{4} }
\institute{Argelander-Institut f\"{u}r Astronomie, Auf dem H\"{u}gel 71, 53121 Bonn, Germany \texttt{nbekhti@astro.uni-bonn.de} \and  Faculty of Information \& Communication Technologies, Swinburne University of Technology, Victoria 3122, Australia \texttt{mmurphy@astro.swin.edu.au} \and Institut f\"{u}r Physik, Am Neuen Palais 10, 14469 Potsdam, Germany \texttt{prichter@astro.physik.uni-potsdam.de} \and Australia Telescope National Facility, Epping NSW 1710, Australia \texttt{tobias.westmeier@csiro.au}  }
\authorrunning{N. Ben Bekhti, M. Murphy, P. Richter, T. Westmeier} 
%
%
\maketitle

Spiral galaxies are surrounded by large gaseous halos that represent the interface between the condensed galactic discs and the surrounding circumgalactic medium.
A powerful method for studying the properties and nature of these circumgalactic gaseous structures is the analysis of absorption-line systems in the spectra of distant quasars (e.g., Richter et al. 2001, Richter et al. 2005). 
We here concentrate on the study of metal-line absorption systems in the halo of our own galaxy and link them with the distribution of neutral gaseous structures in the Milky Way as seen in \ion{H}{i} 21~cm emission.

\begin{figure}[!t]
\centering
\includegraphics[width=5cm,height=6.3cm]{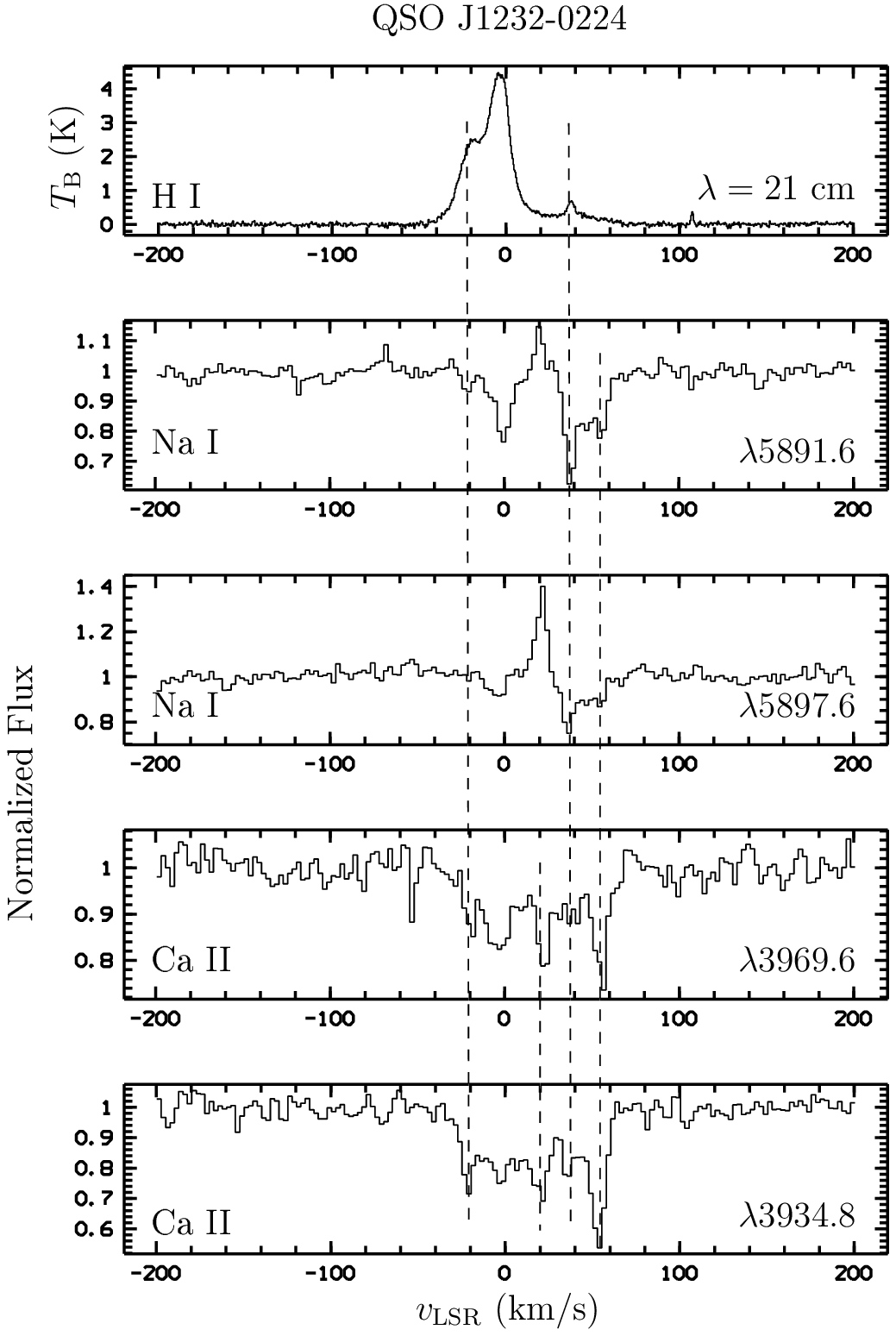}\hfill
\includegraphics[width=6cm]{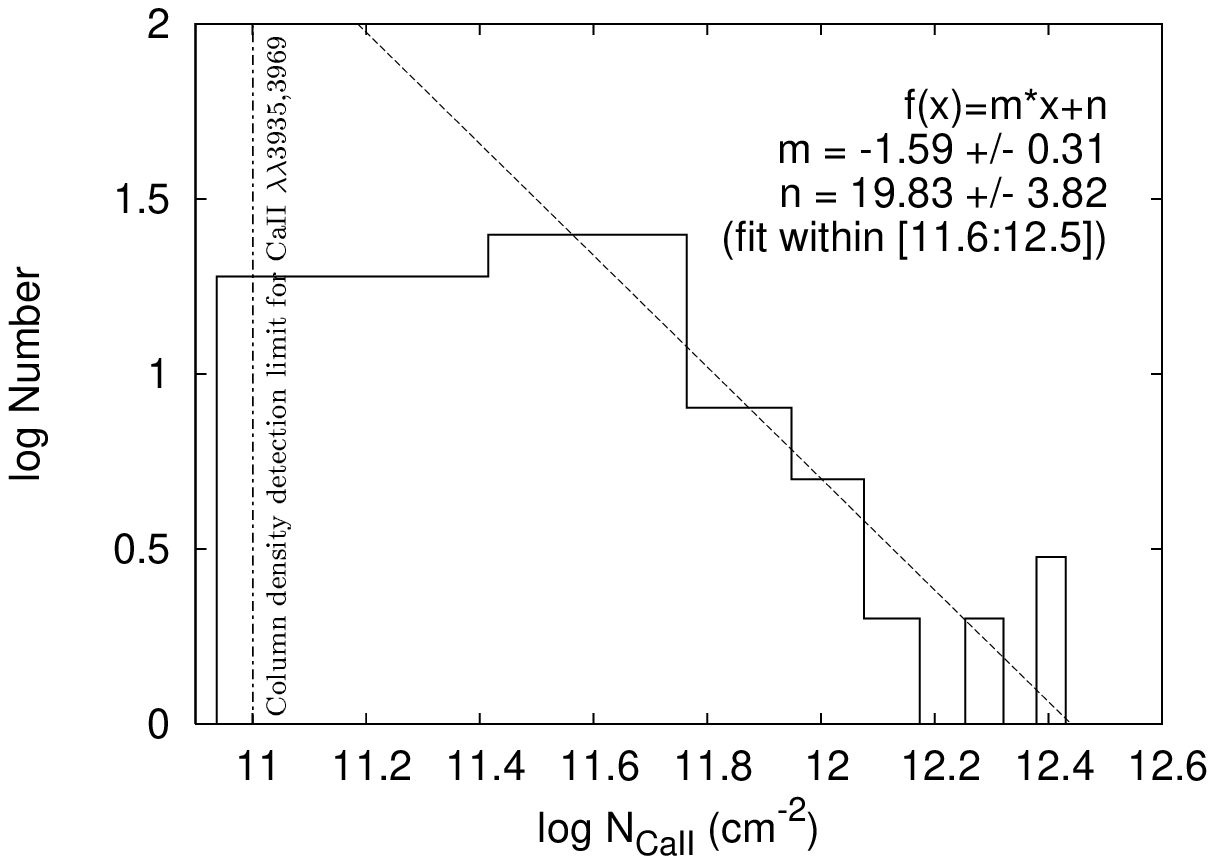}
\caption{\textbf{Left}: \ion{Ca}{ii} and \ion{Na}{i} absorption and the corresponding \ion{H}{i} emission spectra of QSO J1232-0224 obtained with UVES and the Effelsberg 100-m telescope. \textbf{Right}: Logarithm of the number of intermediate- or high velocity absorption lines versus the logarithm of the \ion{Ca}{ii} column density with a linear binning of $\log N_\mathrm{HI}$. The dashed line represent a power law fit with $N^{-1.6}$ defined over the range of column densities from $\log (N_\mathrm{CaII}/\mathrm{cm}^{-2})=11.6$ to $\log (N_\mathrm{CaII}/\mathrm{cm}^{-2})=12.5$.}
\label{fig:1}       
\end{figure}

Our sample contains in total 106 sight lines towards quasars observed in the optical with VLT/UVES. Along 43 of them we
detect halo \ion{Ca}{ii} and/or \ion{Na}{i} absorption lines at intermediate or high velocities. For these we have additional
\ion{H}{i} emission spectra obtained with the 100-m telescope at Effelsberg. The absorption profiles of \ion{Ca}{ii} and
\ion{Na}{i} for one of the sight lines are plotted in the left panel of Fig.~\ref{fig:1} on an LSR velocity scale together
with the corresponding 21-cm emission spectrum. In many cases the \ion{Ca}{ii}/\ion{Na}{i} absorption is connected with
\ion{H}{i} gas at column densities in the range of a few times $10^{18}\,\mathrm{cm}^{-2}$ to $10^{20}\,\mathrm{cm}^{-2}$. Our $3 \sigma$ \ion{H}{i} column density detection limit is about $2 \cdot 10^{18}\,\mathrm{cm}^{-2}$ for the warm neutral medium. In a few cases the detected \ion{H}{i} signals are below the detection limit of large \ion{H}{i} surveys (e.g. LAB survey, Kalberla et al. 2005). The measured \ion{H}{i} line widths imply that for most of the sight lines the gas is relatively cold with temperatures below $1000$\,K. The directions and velocities of several of the clouds suggest a possible association with known HVC or IVC complexes. Most \ion{Ca}{ii}/\ion{Na}{i} absorbers show multiple intermediate- and high-velocity components, indicating the presence of filamentary or clumpy structures. 

The right panel of Fig.~\ref{fig:1} shows the logarithm of the number of sight lines observed with UVES versus $\log
(N_\mathrm{CaII}/\mathrm{cm}^{-2})$. The distribution of the column density follows a power law with slope $\beta=-1.6 \pm
0.3$. The vertical dashed-dotted line indicates the UVES $4 \sigma$ detection limit of $\log
(N_\mathrm{CaII}/\mathrm{cm}^{-2})=11$. Therefore, the lower column density cut-off is only a selection effect determined by
the detection limit of the instrument. As a consequence we are not able to make conclusions about a typical column density
of the \ion{Ca}{ii} absorbers.  

The fact that among 106~random lines of sight through the halo we observe 43~absorption systems with at least one
intermediate- or high-velocity component suggests that the Milky Way halo is filled with low column density gas clouds. Their
investigation with synthesis telescopes will allow us to determine their metallicities and, thus, to learn more about their
origin. Studying the clouds will also allow us to draw conclusions about the absorption line systems commonly observed in the halos of other galaxies (e.g., Bouch{\'e} et al. 2006) and the physical state of this gas.

%
%
%

%
%
%
%
%

%

\begin{thebibliography}{99.}
%
%
%


\bibitem{bouch06} N.~Bouch{\'e}, M.T.~Murphy, C.~P{\'e}roux et al:  MNRAS \textbf{371}, 495 (2006)
\bibitem{kalberla05} P.M.W.~Kalberla, W.B.~Burton, D.~Hartmann et al: A\&A, \textbf{440}, 775 (2005)
\bibitem{richter01} P.~Richter, K.R.~Sembach, B.P.~Wakker et al: ApJ \textbf{562}, L181 (2001)
\bibitem{richter05} P.~Richter, T.~Westmeier, C.~Br\"{u}ns: A\&A \textbf{442}, L49 (2005)


\end{thebibliography}
%



\printindex
\end{document}